\documentclass[twocolumn,showpacs]{revtex4}

\usepackage{graphicx}
\usepackage{dcolumn}
\usepackage{amsmath}
\begin{document}

\title{Quantum oscillations in antiferromagnetic CaFe$_2$As$_2$ on the brink of superconductivity 
}

\author{N.~Harrison$^1$, R.~D.ÊMcDonald$^1$, C.~H.~Mielke$^1$, E.~.D.~Bauer$^1$, F.~Ronning$^1$ and J.~D.~Thompson$^1$
}
\affiliation{
$^1$Los Alamos National Laboratory, MS-E536, Los Alamos, NM 87545\\
}
\date{\today}

\begin{abstract}
We report quantum oscillation measurements on CaFe$_2$As$_2$ under strong magnetic fields$-$ recently reported to become superconducting under pressures of as little as a kilobar. The largest observed carrier pocket occupies less than 0.05~\% of the paramagnetic Brillouin zone volume$-$ consistent with Fermi surface reconstruction caused by antiferromagnetism. On comparing several alkali earth $A$Fe$_2$As$_2$ antiferromagnets (with $A=$~Ca, Sr and Ba), the dependence of both the Fermi surface cross-sectional area $F_\alpha$ and the effective mass $m^\ast_\alpha$ of the primary observed pocket on the antiferromagnetic/structural transition temperature $T_{\rm s}$ is found to be consistent with quasiparticles in a conventional spin-density wave model. These findings suggest that a conventional spin-density wave exists within close proximity to superconductivity in this series of compounds, which may have implications for the microscopic origin of unconventional pair formation.
\end{abstract}
 \maketitle

The detection of magnetic quantum oscillations continues to be a powerful experimental method for determining the existence of Fermi surface reconstruction, often revealing the existence of small pockets of carriers~\cite{doiron1,yelland1,sebastian1} missed by other spectroscopic tools~\cite{hossain1,liu1}. In the recently discovered high $T_{\rm c}$ superconductors based on FeAs layers~\cite{kamihara1, chen1, ren1, liu2, mazin1, takahashi1}, they have played an essential role in establishing the existence of a Fermi surface within the parent antiferromagnetic state~\cite{sebastian1,analytis1}. 
As with cuprates and heavy fermion systems~\cite{monthoux1}, the proximity of antiferromagnetism to superconductivity raises the prospect of antiferromagnetically-mediated superconductivity. 

\begin{figure}
\centering
\includegraphics*[width=0.45\textwidth,angle=0]{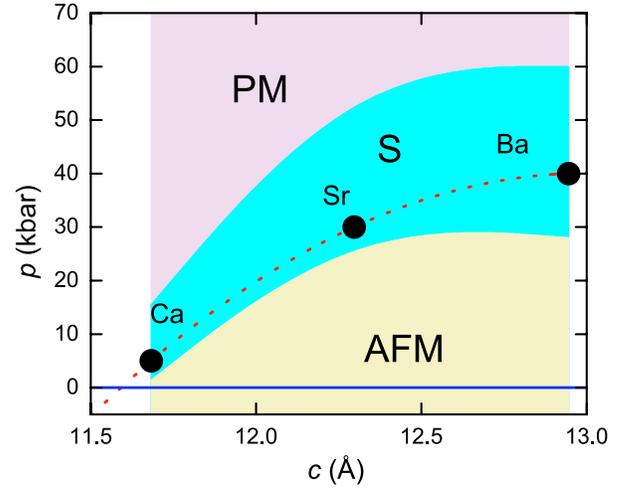}
\caption{Schematic pressure $p$ versus FeAs bilayer spacing $c$ diagram for $A$Fe$_2$As$_2$ where $A=$~Ca, Sr and Ba, indicating the nominal regions where antiferromagnetism (AFM), superconductivity (S) and possible paramagnetic metallic behavior (PM) reside. The black circles (and dotted line) represent the optimal pressure $p_{\rm s}$ at which superconductivity is reported to attain its highest value. The $c$  values appropriate for the orthorhombic antiferromagnetic phase below $T_{\rm s}$ are taken from Refs.~\cite{kreyssig1,tegel1,rotter1}, while the $p$ values are taken from Refs.~\cite{lee1,alireza1}. Lines are drawn between each of the compositions as guides for the eye. 
In CaFe$_2$As$_2$, the sensitivity of the competing antiferromagnetic and `collapsed tetragonal' phases to shear associated with the pressure medium may play some role in fine-tuning superconductivity~\cite{yu1}.
}
\label{spacing}
\end{figure}

Among stoichiometric $A$Fe$_2$As$_2$ compounds, CaFe$_2$As$_2$ lies closest in pressure phase space to superconductivity (see Fig.~\ref{spacing})$-$ reported to exist at pressures as low as 1.4~kbar~\cite{torikachvili1,lee1}, compared to $\sim$~30~kbar in its chemically closest relative SrFe$_2$As$_2$~\cite{alireza1}. One intriguing possibility therefore is that the optimal pressure $p_{\rm s}$ at which superconductivity occurs~\cite{alireza1,torikachvili1,lee1} is correlated with the spacing $c$ between the bilayers (see Fig.~\ref{spacing})~\cite{kreyssig1,tegel1,rotter1}, which is in turn controlled by the size of the alkali earth ion $A$. The consecutive substitution of $A$ with Ca, Sr and Ba thus provides a laboratory for relating possible $c$-dependent changes of the Fermi surface topology and antiferromagnetic correlations with the proximity to pressure-tuned superconductivity.

In the present paper, we determine the Fermi surface of CaFe$_2$As$_2$ by utilizing the contactless conductivity technique that has been successfully applied to SrFe$_2$As$_2$~\cite{sebastian1} and BaFe$_2$As$_2$~\cite{analytis1}, as well as the cuprate superconductors in strong magnetic fields~\cite{yelland1,sebastian2}. A sample of CaFe$_2$As$_2$ of dimensions $\approx$~1~$\times$~1.5~$\times$~0.2~mm$^3$ (preselected to have
minimal Sn flux inclusion) is cut from a larger crystal and mounted with its tetragonal {\it c} axis
aligned parallel to the magnetic induction ${\bf B}\approx\mu_0{\bf
H}$ (or rotated away from ${\bf H}$ by an angle $\theta$). Its planar face is attached to a 5 turn compensated coil of
diameter $\approx$~0.7~mm that forms part of a tunnel diode
oscillator (TDO) circuit. The TDO resonates at a frequency of
$\approx$~46~MHz in the absence of an applied magnetic field,
dropping by $\approx$~200~kHz in $\sim$~60~T in response to the
magnetoresistivity of the sample. The associated increase in in-plane resistivity
changes the coil inductance and resonance frequency, suggesting that the quantum oscillations (which reach a maximum amplitude at $\sim$~60~T of 1 part in a 1000 of the magnetoresistance)
originate from the Shubnikov-de Haas (SdH) effect. 
The TDO frequency is
mixed down to $<$~1~MHz in two stages whereupon it is digitized at a
rate of 20~MHz during the $\sim$~50~ms duration of the magnetic
field pulse. The sample is immersed in gaseous or liquid $^3$He throughout the
experiments, with the temperature controlled through the vapor pressure for $T<$~2K or by a heater for $T>$~2K and measured using a calibrated cernox resistor (mounted close to the sample) immediately prior to the magnetic field pulse.

\begin{figure}
\centering
\includegraphics*[width=0.48\textwidth,angle=0]{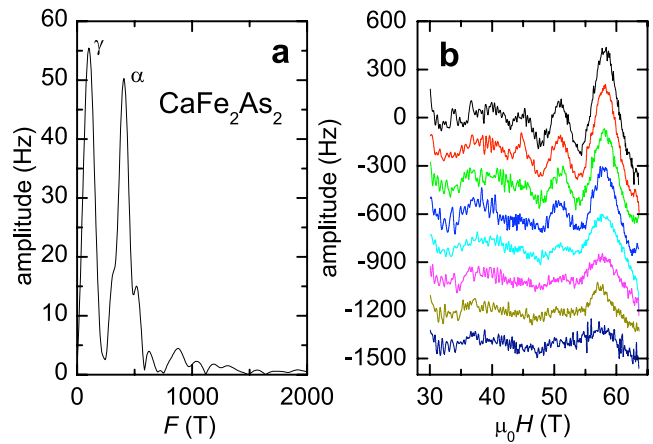}
\caption{{\bf a} Fourier transform of quantum oscillations in the TDO resonance frequency in $1/B$ at 0.98~K. {\bf b} Raw TDO frequency oscillations for ${\bf H}\|{\bf c}$ with curves at $T=$~0.98, 1.52, 2.08, 3.1, 4.2, 5.0, 6.0 and 7.0~K consecutively offset.} \label{data}
\end{figure}

\begin{figure}
\centering
\includegraphics*[width=0.4\textwidth,angle=0]{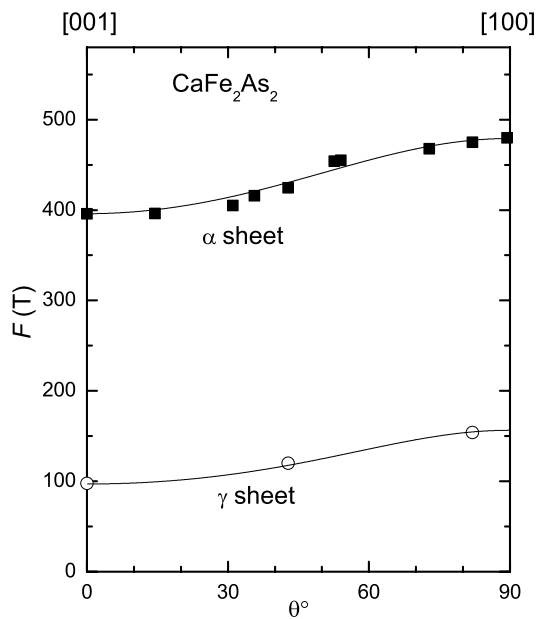}
\caption{Magnetic field orientation-dependence of the $\alpha$ and $\gamma$ quantum oscillation frequencies determined by Fourier transformation, where $\theta=$~0 corresponds to ${\bf H}\|{\bf c}$ and $\theta=$~90$^\circ$ corresponds to ${\bf H}\|{\bf a}$ at room temperature. The sample and coil are rotated in-situ.} \label{angle}
\end{figure}

Figure~\ref{data} shows results of quantum oscillation experiments
performed on CaFe$_2$As$_2$, after subtracting a linear background and performing Fourier transformation in $1/B$. When ${\bf H}\|{\bf c}$, several oscillations are discernable corresponding to a frequency $F_\alpha=$~395~$\pm$~10~T, with another feature at $F_\gamma=$~95~$\pm$~20~T becoming more clearly discernible at higher $T$ when the amplitude of $F_\alpha$ is thermally suppressed (here we use the frequency labeling scheme adopted for SrFe$_2$As$_2$ in Ref.~\cite{sebastian1,note1}). A third frequency $F_\beta\sim$~200~T, if present,  is strongly reduced in amplitude compared to that in SrFe$_2$As$_2$. On further performing a rotation study (presented in Fig.~\ref{angle}), both $F_\alpha$ and $F_\gamma$ are found to be consistent with small ellipsoidal pockets in a similar fashion to SrFe$_2$As$_2$~\cite{sebastian1}, with the larger occupying less than 0.05~\% of the paramagnetic Brillouin zone volume. These findings support a scenario in which the Fermi surface is reconstructed as a consequence of the broken translational symmetry brought about by magnetic ordering, as was recently established in SrFe$_2$As$_2$~\cite{sebastian1}.

\begin{figure}
\centering
\includegraphics*[width=0.4\textwidth,angle=0]{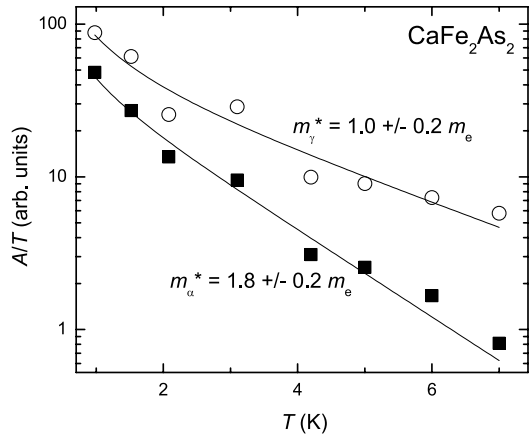}
\caption{Temperature-dependence of $A/T$ for ${\bf H}\|{\bf c}$ together with fits to the Lifshitz-Kosevich theory as described in the text. Fitted values of the effective mass are shown, where $m_{\rm e}$ is the free electron mass.} \label{mass}
\end{figure}

\begin{figure}
\centering
\includegraphics*[width=0.5\textwidth,angle=0]{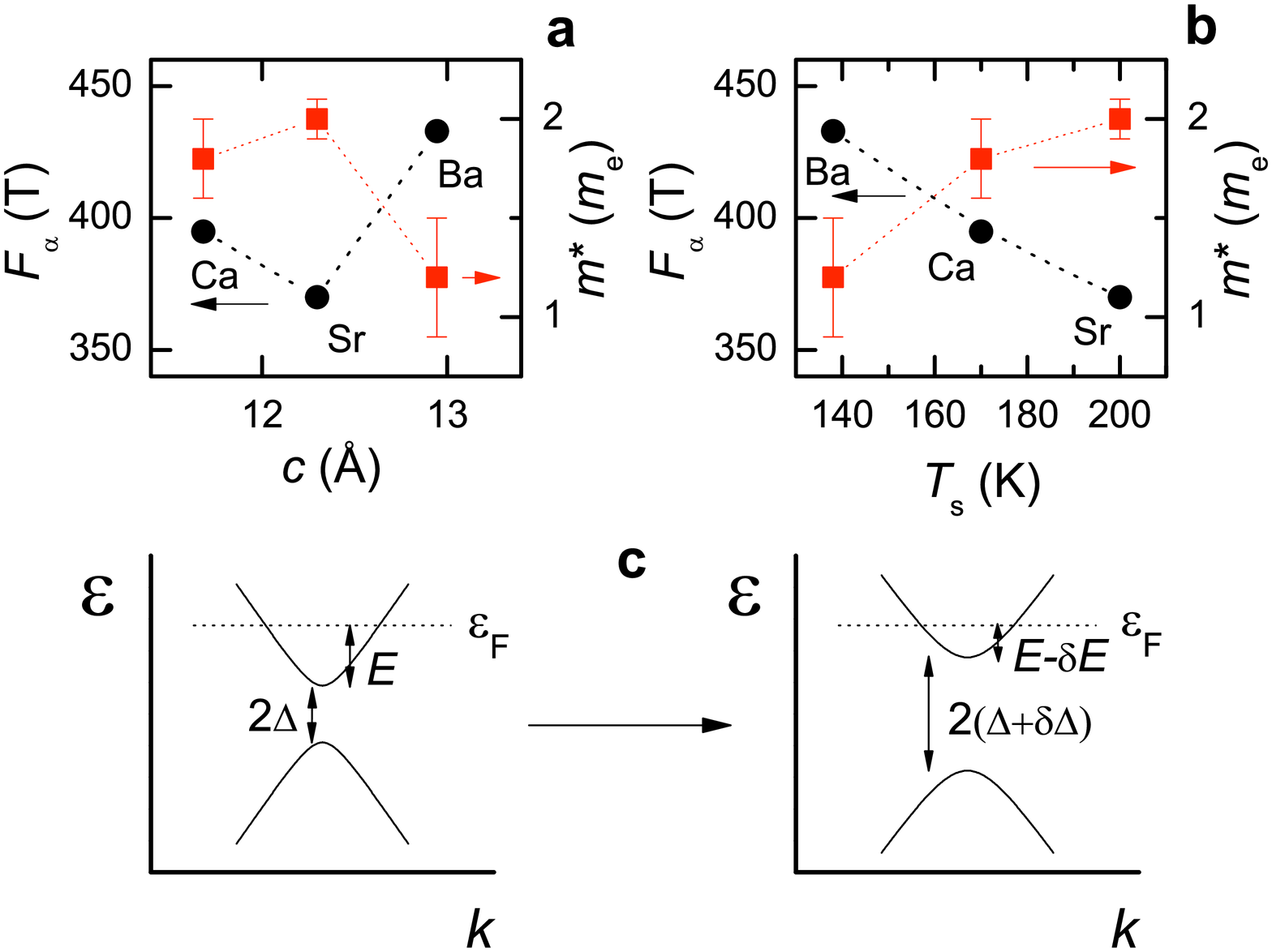}
\caption{Plots of the largest quantum oscillation frequency $F_\alpha$ for ${\bf H}\|{\bf c}$ (right-hand axis) and effective mass $m^\ast_\alpha$ (left-hand-axis) versus the bilayer spacing ({\bf a}) and the structural transition temperature $T_{\rm s}$ ({\bf b}) from the high temperature tetragonal phase into the low temperature orthorhombic, antiferromagnetic phase. {\bf c}, a one-dimensional schematic showing how the size of the pocket ($\propto F$) is inversely related to the size of the spin-density wave gap ($2\Delta\propto T_{\rm s}$). In the simplest model, an increase $\delta\Delta$ in $\Delta$ leads to an equal and opposite reduction $\delta E$ in the filling $E$ of the pocket (for a constant Fermi energy $\varepsilon_{\rm F}$).} \label{SDW}
\end{figure}

A surprising result of the current study, however, is that the effective masses in CaFe$_2$As$_2$ exhibit only a small relative change compared to SrFe$_2$As$_2$ ~\cite{sebastian1} or BaFe$_2$As$_2$~\cite{analytis1}, in spite of CaFe$_2$As$_2$ being in much closer proximity to superconductivity in Fig.~\ref{spacing}. Figure~\ref{mass} shows a fit of the Lifshitz-Kosevich theoretical~\cite{shoenberg1} form $A=A_0X/\sinh{X}$ (where $X=2\pi^2m_ik_{\rm B}T/\hbar eB$) to the temperature-dependent quantum oscillation amplitude in CaFe$_2$As$_2$. The fitted values of the effective mass are compared to those of similar pockets in SrFe$_2$As$_2$ and BaFe$_2$As$_2$ in Table~I and Fig.~\ref{SDW}.


\begin{table}
 \begin{tabular}{ | l | l | l | l | l | l |}
        \hline
        $A$ & $T_{\rm s}$ (K) & $F_\alpha$ (T) & $m^\ast_\alpha$ ($m_{\rm e}$) & $F_\gamma$ (T) & $m^\ast_\gamma$ ($m_{\rm e}$)\\ \hline\hline
        Ca & 170 & 395 & 1.8~$\pm$~0.2 & 95 & 1.0~$\pm$~0.2\\ \hline
        Sr & 200 & 370 & 2.0~$\pm$~0.1 & 70 & $-$\\ \hline
        Ba & 138 & 430 & 1.2~$\pm$~0.3 & 95 & 0.9~$\pm$~0.1\\
        \hline
 \end{tabular}\label{table}
 \caption{Values of the Fermi surface frequencies $F_i$ and effective masses $m_i$ for ${\bf H}\|{\bf c}$, taken from this work and Refs.~\cite{sebastian1,analytis1}. Also shown, is the structural transition temperature $T_{\rm s}$ into the orthorhombic, antiferromagnetic phase.}
 \end{table}
 
The absence of a divergence in the effective mass on reducing $c$ in Fig.~\ref{SDW}a, or reducing the antiferromagnetic/structural transition temperature $T_{\rm s}$  in Fig.~\ref{SDW}b, is in marked contrast to recent observations made in heavy fermion antiferromagnets~\cite{shishido1} and cuprate superconductors~\cite{sebastian3}, where subtle changes in $p$ or doping lead to marked changes in $m^\ast$. Given that the Fermi velocities and.or effective masses are renormalized by only a factor a two compared to bandstructure estimates~\cite{sebastian1,analytis1}, one possibility is that the physics of $A$Fe$_2$As$_2$ is much closer to that of a conventional spin-density wave~\cite{hertz1}, with the on-site Coulomb repulsion between the d-electrons being insufficiently strong to cause development of a Mott insulating state. 
In the case of a conventional spin-density wave, antiferromagnetic correlations lead to a comparatively weak enhancement of $m^\ast$ that increases with the size of the spin-density wave gap $2\Delta$~\cite{mcdonald1}. Differences in the electronic structures of the pnictides, cuprates  and heavy fermions may account for some of their differences in behavior. For example, the electronic structure of FeAs layered compounds differs from the others in that multiple (d-electron) bands cross the Fermi energy in the unreconstructed tetragonal phase~\cite{sebastian1,singh1}. Strong on-site Coulomb repulsion operating within a single d- or f-electron band in heavy fermion and high $T_{\rm c}$ cuprates, by contrast, makes those systems vulnerable to Mott insulating behavior (in which the d- or f-electrons no longer contribute to the electrical conduction)~\cite{si1,imada1}, providing a possible route for the divergence of $m^\ast$ inside their antiferromagnetic states. 

Evidence for conventional spin-density wave ordering in $A$Fe$_2$As$_2$ is provided in Fig.~\ref{SDW}b by the reduction in the size of the pocket $F_\alpha$ and weak enhancement of $m^\ast_\alpha$ on increasing $T_{\rm s}$ (which varies non-monotonically with $A$ and $c$). Here, we concentrate on the larger pocket that is observed consistently in all antiferromagnetic $A$Fe$_2$As$_2$ compounds, which also contributes most to the electronic heat capacity on account of its larger surface area. The observed trends in Fig.~\ref{SDW}b can be understood by considering a simplified spin-density wave reconstruction model (see Fig.~\ref{SDW}c) in which the spin-density wave gap $2\Delta$ is proportional to the transition temperature $T_{\rm s}$ in accordance with the BCS description $\Delta=1.76k_{\rm B}T_{\rm s}$~\cite{gruner1}. 
If the unreconstructed Fermi surfaces are assumed to be similar throughout~\cite{sebastian1}, the progressive increase in $2\Delta$ incurred on going from Ba to Ca to Sr improves the degree of nesting, resulting in a progressive reduction in the sizes of the pockets and a gradual flattening of the reconstructed band leading to the enhancement in $m^\ast$. 

Quantitative consistency with the spin-density wave model can be tested by comparing the reduction $\delta E$ in the filling $E$ of the pocket with the increase $\delta\Delta$ in $\Delta$ on going from $A=$~Ca to Sr (which have the smallest percentage error bars in $m^\ast_\alpha$). If we assume a parabolic dispersion whereby $E=\hbar eF_\alpha/m^\ast_\alpha$ and a BCS gap, we arrive at quantitatively similar incremental changes $\delta E\approx$~4~$\pm$~2~meV and $\delta\Delta\approx$~4.5~meV as expected for the simple spin-density wave model presented in Fig.~\ref{SDW}c.

While spin-density wave physics seems to be relevant to this series of compounds, the process by which it is destabilized under pressure, leading to the emergence of superconductivity, is not yet clearly established. Hydrostatic pressure applied to CaFe$_2$As$_2$ was recently shown to cause a first order transition at $p_{\rm c}\approx$~3.5~kbar into a `collapsed tetragonal phase' with the same crystal symmetry as the room temperature ambient pressure phase but with a greatly reduced lattice spacing $c$~\cite{kreyssig1}. Importantly, the collapsed tetragonal phase has a $\sim$~6~\% reduced volume compared to both the room temperature/ambient pressure tetragonal and orthorhombic phases, and has significantly different lattice parameters.  It was recently argued by Yu {\it et al.}  that by using a pressure medium that solidifies at a higher $T$ than $^4$He, ideal hydrostatic pressure conditions are less likely to be maintained~\cite{yu1}. One possible outcome is that a near-continuous transformation in volume occurs with $p$ under such conditions, leading to phase separation in which orthorhomic and collapsed tetragonal domains coexist, with shear forces enabling both of them to exist out of equilibrium.  It is under these circumstances that superconductivity is reported to exist in CaFe$_2$As$_2$~\cite{torikachvili1,yu1,lee1}. 

While neutron scattering experiments have yet to be performed on SrFe$_2$As$_2$ and BaFe$_2$As$_2$ under high pressures to verify whether the transition into a collapsed tetragonal phase is universal, the proximity to such a phase in CaFe$_2$As$_2$ by virtue of its reduced $c$ may be an important factor in the relative ease by which superconductivity can be tuned~\cite{kreyssig1}. The sensitivity of the superconducting volume fraction in SrFe$_2$As$_2$ and BaFe$_2$As$_2$ to pressure provides some hint of a phase separation scenario analogous to that in CaFe$_2$As$_2$~\cite{sebastian1}. Identifying which of the phases is primarily responsible for the observed superconductivity in CaFe$_2$As$_2$ and the directionality of the shear required for its optimization remains a formidable challenge.

The discontinuous nature of the transition into the collapsed tetragonal phase with a greatly reduced bilayer spacing compared to the FeAs-layered compounds with higher $T_{\rm c}$'s, suggests that the transition itself is unlikely to be a source of diverging antiferromagnetic fluctuations. An alternative possibility is that superconductivity emerges from within the orthorhombic phase by way of a continuous suppression of antiferromagnetism by shear-restored geometric frustration of the staggered Fe moments or by disorder. This would yield a quantum critical point qualitatively different~\cite{si2} from that usually associated with pressure-tuned superconductivity in heavy fermion compounds~\cite{shishido1,hertz1,si1}. A large spacing between the layers is preserved in the orthorhombic phase, which appears to be an important factor in attaining high superconducting transition temperatures in FeAs layered compounds~\cite{rotter2}, as it is in other layered superconductors~\cite{schrieffer1,bauer1}. 

In summary, while there remains uncertainty over the precise conditions required to optimize superconductivity with pressure, CaFe$_2$As$_2$ is arguably the stoichiometric antiferromagnet existing within closest proximity to superconductivity. A comparison of the Fermi surface properties of CaFe$_2$As$_2$ with those of SrFe$_2$As$_2$ and BaFe$_2$As$_2$ reveals that the effective masses and pocket sizes exhibit a dependence on the antiferromagnetic/structural transition temperature $T_{\rm s}$, that is consistent with weakly enhanced quasiparticles moving in an antiferromagnetic background of spins. These findings suggest that antiferromagnetism in $A$Fe$_2$As$_2$ can be effectively understood from the perspective of a conventional spin-density wave$-$ likely placing constraints on possible mechanisms for superconductivity in FeAs layered compounds~\cite{monthoux1,si1,si2}.

This work is conduced under the auspices the US Department of Energy, while the National High Magnetic Field Laboratory, where the experiments were conducted, is primarily funded by the National Science Foundation and the State of Florida.


\begin{thebibliography}{99}
\bibitem{note1} Note that the $\alpha$ and $\beta$ are mistakenly interchanged in Fig.~2b of Ref.~\cite{sebastian1} and in the discussion section of Ref.~\cite{sebastian1} relating the pockets to the bandstructure calculations.
\end{thebibliography}
\end{document}